# Some properties of coherent states with singular complex matrix argument

Dušan POPOV[a, b]

[a] University Politehnica Timisoara, Department of Physical Foundations of Engineering, B-dul Vasile Pârvan No. 2, 300223 Timisoara, Romania
[b] Serbian Academy of Nonlinear Sciences (SANS), Kneza Mihaila 36, Beograd-Stari Grad, Belgrade, Serbia
E-mail: dusan_popov@yahoo.co.uk
ORCID: https://orcid.org/0000-0003-3631-3247

**Abstract**

In the paper our aim was to study the properties of a new version of coherent states whose argument is a linear combination of two special singular square 2 x 2 matrix, having a single nonzero element, equal to 1, and two labeling complex variables as developing coefficients. We have shown that this new version of coherent states satisfies all the conditions imposed on coherent states, both of pure, as well as the mixed (thermal) states characterized by the density operator. As applications, we examined the connection between these coherent states and the notions of qubits and von Neuman entropy.

**1. Introduction**

Coherent states, discovered exactly a century ago (Schrodinger, 1926, although he did not use this name) [1], have proven to be a useful approach in various branches of theoretical physics, especially in quantum optics and quantum information. From the first attempt, referring to the linear quantum harmonic oscillator, over time coherent states were formulated for more complex quantum systems, respectively for nonlinear oscillators. The methods for describing the properties of these states were diverse, connected with phase space considerations, operator algebra, semi classical approximations, geometric quantization and so on ([2], [3], [4], [5], [6], and the references therein). Among the newest methods, one can consider the use of bicomplex algebra which highlights the properties of bicomplex numbers. Namely, a new generation of coherent states has been built, based on the use of the generalized hypergeometric functions which have bicomplex arguments [7], [8].

From another point of view, another relatively new direction of examining the properties of coherent states was the one in which the argument of these states is of matrix type, i.e. the states in which the parameter which label the coherent states was replaced by an $n \times n$ valued function [9], [10].

Coherent states with shifted argument, as well as the displaced number states are also starting to attract attention lately [11], [12], [13].

On the other hand, if we consider a vector space of the square diagonal matrix of dimensions $2 \times 2$, then it can define a so called Hamel basis [14]. This means that any matrix can be uniquely written as a finite linear combination of the standard Hamel basis matrix, i.e. the set of special singular unitary matrix (which have only one element equal with 1, the rest being zero).

$$\begin{pmatrix} a & b \\ c & d \end{pmatrix} = a \begin{pmatrix} 1 & 0 \\ 0 & 0 \end{pmatrix} + b \begin{pmatrix} 0 & 1 \\ 0 & 0 \end{pmatrix} + c \begin{pmatrix} 0 & 0 \\ 1 & 0 \end{pmatrix} + d \begin{pmatrix} 0 & 0 \\ 0 & 1 \end{pmatrix} \quad (1.1)$$

where the matrix elements $a, b, c, d$ are real or complex numbers.

Generally, by definition, a matrix is singular if its determinant is equal to zero. Consequently, this type of matrix is non-invertible. For the case of diagonal square matrices, in order to be singular, it is sufficient for one of the diagonal elements to be equal to zero.

In the present paper we aim to examine the properties of a particular case of singular diagonal matrix, which has on the main diagonal a single nonzero element, equal to 1, the rest of the elements being equal to zero.

We will examine such a type of coherent states, labeled by a linear combination of two singular diagonal matrices of dimensions 2 x 2, which have as expansion coefficients two complex variables.

Therefore, we will consider the following two basic singular matrices:

$$u_0 \equiv \begin{pmatrix} 1 & 0 \\ 0 & 0 \end{pmatrix} \quad , \quad u_1 \equiv \begin{pmatrix} 0 & 0 \\ 0 & 1 \end{pmatrix} \quad (1.2)$$

In other words, a $2 \times 2$ matrix, as an element of the vector space, i. e. a diagonal square matrix $Z$, can be written as a linear combination of Hamel type, so that the coefficients are complex variables. It can be represented as follows

$$Z \equiv \begin{pmatrix} z & 0 \\ 0 & \sigma \end{pmatrix} = z\, u_0 + \sigma\, u_1 = z \begin{pmatrix} 1 & 0 \\ 0 & 0 \end{pmatrix} + \sigma \begin{pmatrix} 0 & 0 \\ 0 & 1 \end{pmatrix} \quad (1.3)$$

Consequently, for the two dimensional case, which we will examine, the argument we will use is $Z = z\, u_0 + \sigma\, u_1$. This leads to functions that satisfy Cauchy's functional equation [15], but which involve complex variables

$$\mathcal{F}(z\, u_0 + \sigma\, u_1) = \mathcal{F}(z\, u_0) + \mathcal{F}(\sigma\, u_1) = \mathcal{F}(z)u_0 + \mathcal{F}(\sigma)u_1 \quad (1.4)$$

In this idea, in the present paper we propose to study the properties of coherent states examined from two different angles: on the one hand as coherent states with shifted complex argument, and on the other hand as coherent states with linear combination argument of special 2 x 2 singular matrices, involving the Hamel basis. This is the main purpose of the present paper. Due to the properties of these matrices, functions labeled by two complex variables will decompose into two functions depending on only one variable each. Our 2 x 2 matrices are somewhat similar to a quaternion matrix (i.e. an extension of complex numbers formed by a real number and a three-dimensional vector), but the matrices we use have specific properties.

As a side note, let us point out here that complex representations of quaternions have been used to construct the so-called coherent vector states [9], [10], [11], [16], [17].

## 2. Some properties of special square orthogonal matrices

As is known, any quantum state of a system is represented by a linear combination of state vectors, the set of these vectors forming an orthonormal basis in Hilbert space (the vectors being mutually perpendicular and of norm equal to one). One of the most useful bases is the Fock vector basis, which uses states with occupation number. Each individual state is the state of a single particle, being itself the eigenvalue of the particle number operator, orthogonal and forming a complete set for representing any arbitrary quantum state in Fock space. If we use the matrix form, each Fock vector is represented by a column matrix, in which all terms are zero, except for one which is equal to 1 and which represents the state containing a single particle.

So, the column vector $|n>$, of dimensions $n_{max} \times 1$ and its transpose (row vector) are $<n|=|n>^T$ with dimensions $1 \times n_{max}$ have all elements equal to zero, except for the element in place $n+1$, which is equal to 1 :

$$\mathcal{M}(|n>)_{n_{max} \times 1} \equiv (\delta_{i1})_{n_{max} \times 1} \; , \; \delta_{i1} = \begin{cases} 0, & i \neq n \\ 1, & i = n \end{cases} \; , \; \mathcal{M}(<n|)_{1 \times n_{max}} \equiv (\delta_{1j})_{1 \times n_{max}} \; , \; \delta_{1j} = \begin{cases} 0, & j \neq n \\ 1, & j = n \end{cases} \qquad (2.1)$$

$$\mathcal{M}(<n|)_{1 \times n_{max}} \cdot \mathcal{M}(|n>)_{n_{max} \times 1} = 1 \qquad (2.2)$$

In expanded form, the Fock vectors are expressed as follows

$$|0> \equiv \begin{pmatrix} 1 \\ 0 \\ \vdots \\ 0 \end{pmatrix} \; , \; |1> \equiv \begin{pmatrix} 0 \\ 1 \\ \vdots \\ 0 \end{pmatrix} \; , \; ..., \; |n_{max}> \equiv \begin{pmatrix} 0 \\ 0 \\ \vdots \\ 1 \end{pmatrix} \qquad (2.3)$$

Consequently, the Fock's vectors projector $|n><n|$ is a square singular diagonal matrix of dimensions $n_{max} \times n_{max}$, which has only one element equal to 1 on the main diagonal, the remaining off-diagonal elements being equal to zero. We will denote the projector by $u_n$:

$$u_n \equiv |n><n| = \mathcal{M}(|n>)_{n_{max} \times 1} \cdot \mathcal{M}(<n|)_{1 \times n_{max}} \equiv (\delta_{ij})_{n_{max} \times n_{max}} \; , \; \delta_{ij} = \begin{cases} 0, & i,j \neq n \\ 1, & i,j = n \end{cases} \qquad (2.4)$$

It is observed that $u_j$, $j = 0, 1, ..., n_{max}$ matrices have a special structure: they have a single element, on the main diagonal, equal to 1, all other elements being equal to zero. It can be verified that these matrices have the following properties: they are positive, semi-definite, idempotent and mutually orthogonal. Their determinant is zero and for this reason the matrices are singular and non-invertible. Because they are diagonal, as a consequence, there are always symmetric. The product of any two different matrix is the zero matrix. The trace is equal to 1.

$$u_i(u_j)^m = (u_i)^m u_j = \begin{cases} u_j, & \text{for } i = j \\ 0, & \text{for } i \neq j \end{cases}, \quad i,j = 0, 1, ..., n_{max}, \quad m = 1, 2, ... \qquad (2.5)$$

$$\sum_{n=0}^{n_{max}-1} u_n = \sum_{n=0}^{n_{max}-1} |n><n| = I_{n_{max}} \qquad (2.6)$$

As an example which is often used in quantum mechanics, let us we consider a system with $n_{max}$ states. If the system is in thermodynamic equilibrium with the environment, at the temperature $T = 1/\beta k_B$, then their states are mixed and described by the normalized canonical density operator. It can be represented as a linear combination of diagonal square matrices, having only one nonzero element on the main diagonal, equal to 1.

$$\rho = \frac{1}{Z(\beta)} \sum_{n=0}^{n_{max}-1} e^{-\beta E_n} |n><n| = \frac{1}{Z(\beta)} \sum_{n=0}^{n_{max}-1} e^{-\beta E_n} u_n, \quad Z(\beta) = \sum_{n=0}^{n_{max}-1} e^{-\beta E_n} \qquad (2.7)$$

Hence the interest in this type of matrices and, at the same time, the motivation to examine their properties in connection with applications in quantum mechanics, especially in coherent states approach. We will limit ourselves (and for reasons of description of the formulas - shorter) only to 2 x 2 matrices, although the results obtained can be generalized relatively easily to the case of matrices of dimensions $n_{max} \times n_{max}$.

Particularly, let us we consider a two-level energy system, where $E_0$ is the energy eigenvalue in the ground state $n = 0$, respectively $E_1$ - in the excited state, $n = 1$. The corresponding density operator in $C^2$ has the following spectral decomposition

$$\rho = \frac{1}{Z(\beta)} \sum_{n=0}^{1} e^{-\beta E_n} |n><n| = \frac{1}{Z(\beta)} e^{-\beta E_0} |0><0| + \frac{1}{Z(\beta)} e^{-\beta E_1} |1><1| \qquad (2.8)$$

with the partition function (the sum of diagonal elements)

$$Z(\beta) = \text{Tr}\,\rho = e^{-\beta E_0} + e^{-\beta E_1} \qquad (2.9)$$

In matrix manner, the two energy states are

$$|0> = \begin{pmatrix} 1 \\ 0 \end{pmatrix}, \quad <0| = (1 \; 0), \quad |0><0| = \begin{pmatrix} 1 \\ 0 \end{pmatrix} \cdot (0 \; 1) = \begin{pmatrix} 1 & 0 \\ 0 & 0 \end{pmatrix} \equiv u_0 \qquad (2.10)$$

$$|1> = \begin{pmatrix} 0 \\ 1 \end{pmatrix}, \quad <1| = (0 \; 1), \quad |1><1| = \begin{pmatrix} 0 \\ 1 \end{pmatrix} \cdot (0 \; 1) = \begin{pmatrix} 0 & 0 \\ 0 & 1 \end{pmatrix} \equiv u_1 \qquad (2.11)$$

In quantum mechanics, when an observable has a spectrum of discrete eigenvalues, the set $\{|n>\}$ being complete, any ket $|\Psi>$ can be uniquely expanded as the linear combination, the sum extending over the entire range of values for $n$, i.e. $|\Psi> = \sum_n c_n |n>$. The same situation occurs in the case of singular diagonal square matrices, which can be written as a linear combination of column matrices, which have all zero elements, except for one, equal to 1.

$$\mathcal{M}(z_n)_{n_{max} \times n_{max}} \equiv (z_n \delta_{ij}) \quad , \quad \delta_{ij} = \begin{cases} 0, & i,j \neq n \\ z_n, & i,j = n \end{cases} \quad , \quad n = 0,1,...,n_{max} \tag{2.12}$$

$$\mathcal{M}(z_n)_{n_{max} \times n_{max}} = \sum_{n=0}^{n_{max}-1} z_n \, u_n \tag{2.13}$$

This is in fact the linear expansion in the Hamel basis [14] (that is a maximal linearly independent set spanning a vector space, using finite combinations) of a set of linearly independent diagonal square matrices (vectors) covering the entire space of this type of matrix. The expansion is unique and the coefficients of the expansion are the complex numbers, i.e. the elements of the singular diagonal matrix.

Let's recall the properties of these matrices:

$$(u_0)^2 = u_0 \, , \, (u_1)^2 = u_1 \, , \, u_i u_j = \begin{cases} 1, & \text{for } i = j \\ 0, & \text{for } i \neq j \end{cases} \, , \, i,j = 0,1 \, , \, u_0 + u_1 = \begin{pmatrix} 1 & 0 \\ 0 & 1 \end{pmatrix} \equiv I \tag{2.14}$$

As geometrical interpretations, the matrices $u_j$ are in fact the orthogonal projection matrices which project vectors onto the real coordinate axes, $u_0 \equiv P_x$ and $u_1 \equiv P_y$:

$$P_x \begin{pmatrix} x \\ y \end{pmatrix} = \begin{pmatrix} x \\ 0 \end{pmatrix} \quad , \quad P_y \begin{pmatrix} x \\ y \end{pmatrix} = \begin{pmatrix} 0 \\ y \end{pmatrix} \tag{2.15}$$

These projection matrices represent the probability density that a system that has only two states (or two energy levels) will end up in one of these states.

And, finally, any 2 x 2 diagonal matrix can be written as a linear combination of these two projective matrices.

$$\begin{pmatrix} z & 0 \\ 0 & \sigma \end{pmatrix} = z \begin{pmatrix} 1 & 0 \\ 0 & 0 \end{pmatrix} + \sigma \begin{pmatrix} 0 & 0 \\ 0 & 1 \end{pmatrix} = z \, u_0 + \sigma \, u_1 \tag{2.16}$$

This can be related to Cauchy's functional equation

$$\mathcal{F}(z \, u_0 + \sigma \, u_1) = \mathcal{F}(z \, u_0) + \mathcal{F}(\sigma \, u_1) \tag{2.17}$$

Particularly, an important property of these 2 x 2 diagonal matrices, which we will make full use of in the calculations that follow, is:

$$\mathcal{F}(x \, u_j) = \mathcal{F}(x) u_j \quad , \quad j = 0, 1 \tag{2.18}$$

The proof is immediate:

$$\mathcal{F}(xu_j) = \sum_{n=0}^{\infty} c_n (xu_j)^n = \sum_{n=0}^{\infty} c_n x^n u_j^n = \sum_{n=0}^{\infty} c_n x^n u_j = u_j \sum_{n=0}^{\infty} c_n x^n = \mathcal{F}(x) u_j \quad (2.19)$$

This means that if the matrix $u_j$ appears as a factor in the argument of an analytic function, then he can be extracted from this function.

$$\mathcal{F}(zu_0 + \sigma u_1) \equiv \mathcal{F}\left(\begin{pmatrix} z & 0 \\ 0 & \sigma \end{pmatrix}\right) = \mathcal{F}(z)u_0 + \mathcal{F}(\sigma)u_1 = \begin{pmatrix} \mathcal{F}(z) & 0 \\ 0 & \mathcal{F}(\sigma) \end{pmatrix} \quad (2.20)$$

In this case, the Cauchy's functional equation will be written as

$$\mathcal{F}(xu_0 + yu_1) = \mathcal{F}(x)u_0 + \mathcal{F}(y)u_1 \quad (2.21)$$

Returning to the density operator of a two-level system, with the new notations we will be able to write:

$$\rho = \frac{1}{Z(\beta)} \sum_{n=0}^{1} e^{-\beta E_n} |n\rangle\langle n| = \frac{1}{Z(\beta)} \left( e^{-\beta E_0} u_0 + e^{-\beta E_1} u_1 \right) = \begin{pmatrix} e^{-\beta E_0} & \\ & e^{-\beta E_1} \end{pmatrix} \quad (2.22)$$

The partition function $Z(\beta)$ is obtained from the normalization condition of the density operator, $\text{Tr}\rho = 1$:

$$Z(\beta) = \sum_{m=0}^{1} \langle m|\rho|m\rangle = \sum_{m=0}^{1} e^{-\beta E_m} = e^{-\beta E_0} + e^{-\beta E_1} \quad (2.23)$$

For this case, the density operator is a 2 x 2 diagonal matrix

$$\rho = \frac{1}{1 + e^{-\beta(E_1 - E_0)}} u_0 + \frac{1}{1 + e^{+\beta(E_1 - E_0)}} u_1 = \begin{pmatrix} \frac{1}{1 + e^{-\beta(E_1 - E_0)}} & 0 \\ 0 & \frac{1}{1 + e^{+\beta(E_1 - E_0)}} \end{pmatrix} \quad (2.24)$$

from which it can be seen that the trace of this operator is equal to unity

$$\text{Tr}\rho = \frac{1}{1 + e^{-\beta(E_1 - E_0)}} + \frac{1}{1 + e^{+\beta(E_1 - E_0)}} = 1 \quad (2.25)$$

### 3. Coherent states with shifted arguments

In a previous paper [13] we have deduced the expression of generalized coherent states, with a displaced or shift argument. This type of coherent states has as its argument a linear combination of two complex variables $z$ and $\sigma$, and two real numbers $\varepsilon$ and $\lambda$ as coefficients. Let us point out below the main results regarding coherent states with shifted argument, which we obtained in this paper. These results will be useful to us in examining the topic we are now dealing with.

We used a pair of following nonlinear adjoint annihilation / creation operators

$$\hat{\mathcal{A}}_{-} = \hat{a} \,_p f_q(\hat{\mathcal{N}}) \quad , \quad \hat{\mathcal{A}}_{+} = \,_p f_q(\hat{\mathcal{N}}) \hat{a}^{+} \tag{3.1}$$

with $_p f_q(\hat{\mathcal{N}})$ the deformation function depending on the number operator $\hat{\mathcal{N}} = \hat{a}^{+}\hat{a}$, and $p$ and $q$ positive integers. The deformation function change the usual standard canonical commutation relation $[\hat{a}, \hat{a}^{+}] = 1$ into new deformed Heisenberg algebra. It satisfy the following properties

$$_p f_q(\hat{\mathcal{N}} + m)|n+s\rangle = \,_p f_q(n+s+m)|n+s\rangle \tag{3.2}$$

$$_p f_q(\hat{\mathcal{N}}) = \sqrt{\frac{\prod_{j=1}^{q}(b_j - 1 + \hat{\mathcal{N}})}{\prod_{i=1}^{p}(a_i - 1 + \hat{\mathcal{N}})}} \quad , \quad _p f_q(\hat{\mathcal{N}})|n\rangle = \sqrt{\frac{\prod_{j=1}^{q}(b_j - 1 + n)}{\prod_{i=1}^{p}(a_i - 1 + n)}}|n\rangle \tag{3.3}$$

The actions of these operators and their adjoint on the Fock vectors are

$$\hat{\mathcal{A}}_{-}|n\rangle = \sqrt{_p e_q(n)}|n-1\rangle \quad , \quad \langle n|\hat{\mathcal{A}}_{+} = \sqrt{_p e_q(n)}\langle n-1|,$$
$$\hat{\mathcal{A}}_{+}|n\rangle = \sqrt{_p e_q(n+1)}|n+1\rangle , \quad \langle n|\hat{\mathcal{A}}_{-} = \sqrt{_p e_q(n+1)}\langle n+1| \tag{3.4}$$
$$\langle n|\hat{\mathcal{A}}_{+}\hat{\mathcal{A}}_{-}|n\rangle = \,_p e_q(n)$$

were we used the notation

$$_p e_q(n) \equiv n[_p f_q(n)]^2 = n\frac{\prod_{j=1}^{q}(b_j - 1 + n)}{\prod_{i=1}^{p}(a_i - 1 + n)} \tag{3.5}$$

The repeated action of the creation operator on the vacuum state leads to

$$(\hat{\mathcal{A}}_{+})^n|0\rangle = \sqrt{\prod_{s=1}^{n} {}_p e_q(s)}|n\rangle \equiv \sqrt{_p \rho_q(n)}|n\rangle$$
$$\langle 0|(\hat{\mathcal{A}}_{-})^n = \langle n|\sqrt{\prod_{s=1}^{n} {}_p e_q(s)} \equiv \langle n|\sqrt{_p \rho_q(n)} \tag{3.6}$$

The newly appeared quantity is called the structure function because it determines the internal structure of coherent states.

$$_p \rho_q(n) = \prod_{s=1}^{n} {}_p e_q(s) = n!\frac{\prod_{j=1}^{q}(b_j)_n}{\prod_{i=1}^{p}(a_i)_n} = \Gamma(\boldsymbol{a}/\boldsymbol{b})\Gamma(n+1)\frac{\prod_{j=1}^{q}\Gamma(b_j + n)}{\prod_{i=1}^{p}\Gamma(a_i + n)} \quad , \quad \Gamma(\boldsymbol{a}/\boldsymbol{b}) = \frac{\prod_{i=1}^{p}\Gamma(a_i)}{\prod_{j=1}^{q}\Gamma(b_j)} \tag{3.7}$$

Instead of the usual binomial coefficients in calculations, generalized binomial coefficients appear, defined as follows:

$$\binom{_p\rho_q(l)}{_p\rho_q(n)} = \frac{_p\rho_q(l)}{_p\rho_q(n)\,_p\rho_q(l-n)} \tag{3.8}$$

The notion of generalized binomial coefficient intervenes in the calculations for finding the expression of coherent states with shifted argument, where a product of two shift operators is applied to the vacuum state, each depending on a single complex variable [13]. Also, Newton's generalized binomial expression is used in the calculations, a quantity noted between square brackets $[x+y]^l$ and defined as

$$[x+y]^l \equiv \sum_{m=0}^{l} \binom{\rho(l)}{\rho(m)} x^{l-m} y^m \tag{3.9}$$

The coherent states with shifted argument were labeled by a complex number $Z$ which is a linear combination of two ordinary complex variables, $z$ and $\sigma$, and $\varepsilon$ and $\lambda$ as real numbers:

$$Z = \varepsilon z + \lambda \sigma = \mathcal{R}e\, Z + i\, \mathcal{I}m\, Z = |Z|\exp(i\varphi_Z) \tag{3.10}$$

In deducing the expression of generalized coherent states with shifted arguments, in the paper [13] we used the generalized displacement operator which has the expression (this agrees with [11]):

$$\hat{\mathcal{D}}(z) \equiv \frac{1}{\sqrt{_pF_q(|z|^2)}}\,_pF_q(z\hat{\mathcal{A}}_+) \tag{3.11}$$

To simplify the writing of formulas, in the following, where there might be no confusion, we will omit writing the sets of numbers $\boldsymbol{a} \equiv \{a_i\}_1^p$ and $\boldsymbol{b} \equiv \{b_j\}_1^q$ from the expression of generalized hypergeometric functions. So we will just write $_pF_q(x)$, instead of $_pF_q(\boldsymbol{a};\boldsymbol{b};x)$, i.e.

$$_pF_q(\boldsymbol{a};\boldsymbol{b};x) \equiv\, _pF_q(x) = \sum_{n=0}^{\infty} \frac{\prod_{i=1}^{p}(a_i)_n}{\prod_{j=1}^{q}(b_j)_n} \frac{x^n}{n!} = \sum_{n=0}^{\infty} \frac{x^n}{_p\rho_q(n)} \tag{3.12}$$

Consequently, coherent states having as argument the linear combination of complex numbers is defined using the generalized displacement operator $\hat{\mathcal{D}}(\varepsilon z + \lambda \sigma) = \hat{\mathcal{D}}(\lambda \sigma)\hat{\mathcal{D}}(\varepsilon z)$ which act on the vacuum state:

$$|Z> \equiv |\varepsilon z + \lambda \sigma > = \hat{\mathcal{D}}(\lambda \sigma)\hat{\mathcal{D}}(\varepsilon z)|0> \tag{3.13}$$

After simple calculations, the final expression for displaced (shifted) argument was obtained as

$$|\varepsilon z + \lambda \sigma > = \sqrt{\frac{_pF_q(|\lambda\sigma|^2)\,_pF_q(|\varepsilon z|^2)}{_pF_q(|[\varepsilon z + \lambda \sigma]|^2)}}\,\hat{\mathcal{D}}(\lambda\sigma)\hat{\mathcal{D}}(\varepsilon z)|0> \tag{3.14}$$

or, in equivalent manner, as their expansion with respect to the Fock vectors

$$|\varepsilon z+\lambda\sigma>=\frac{1}{\sqrt{{}_{p}F_{q}(|[\varepsilon z+\lambda\sigma]|^{2})}}\sum_{n=0}^{\infty}\frac{[\varepsilon z+\lambda\sigma]^{n}}{\sqrt{{}_{p}\rho_{q}(n)}}|n> \qquad (3.15)$$

This can also be written in the following form:

$$|\varepsilon z+\lambda\sigma>=\frac{1}{\sqrt{{}_{p}F_{q}(|[\varepsilon z+\lambda\sigma]|^{2})}}{}_{p}F_{q}([\varepsilon z+\lambda\sigma]\hat{\mathcal{A}}_{+})|0> \qquad (3.16)$$

and similar for their complex conjugate counterpart.

Using the variable $Z=\varepsilon z+\lambda\sigma$, the coherent states will be

$$|Z>=\frac{1}{\sqrt{{}_{p}F_{q}(|[Z]|^{2})}}\sum_{n=0}^{\infty}\frac{[Z]^{n}}{\sqrt{{}_{p}\rho_{q}(n)}}|n>=\frac{1}{\sqrt{{}_{p}F_{q}(|[Z]|^{2})}}{}_{p}F_{q}([Z]\hat{\mathcal{A}}_{+})|0> \qquad (3.17)$$

$$<Z|=\frac{1}{\sqrt{{}_{p}F_{q}(|[Z]|^{2})}}\sum_{n=0}^{\infty}\frac{[Z^{*}]^{n}}{\sqrt{{}_{p}\rho_{q}(n)}}<n|=\frac{1}{\sqrt{{}_{p}F_{q}(|[Z]|^{2})}}<0|{}_{p}F_{q}([Z^{*}]\hat{\mathcal{A}}_{+}) \qquad (3.18)$$

### 4. Coherent states with matrix arguments

Let's consider a complex variable $Z$ that has a diagonal matrix structure, i.e. is a sum of two complex numbers $z$ and $\sigma$, each multiplied by a 2 x 2 matrix:

$$Z=\begin{pmatrix} z & 0 \\ 0 & \sigma \end{pmatrix}=z\begin{pmatrix} 1 & 0 \\ 0 & 0 \end{pmatrix}+\sigma\begin{pmatrix} 0 & 0 \\ 0 & 1 \end{pmatrix}=z\,u_{0}+\sigma\,u_{1} \qquad (4.1)$$

Therefore, the complex number $Z$ is a 2x2 diagonal matrix, in which the elements on the main diagonal are the ordinary complex numbers $z$ and $\sigma$, which at the same time are also coefficients of a linear combination of two singular diagonal square matrices $u_{j}$, $j=0,1$.

Using the results obtained for coherent states with shifted argument, let us examine the corresponding properties of coherent states whose argument is a linear combination of complex numbers and two singular square 2 x 2 diagonal matrices.

First let's make the following identifications:

$$\varepsilon z+\lambda\sigma\equiv z\,u_{0}+\sigma\,u_{1}, \quad \varepsilon\equiv u_{0}=\begin{pmatrix} 1 & 0 \\ 0 & 0 \end{pmatrix}, \quad \lambda\equiv u_{1}=\begin{pmatrix} 0 & 0 \\ 0 & 1 \end{pmatrix} \qquad (4.2)$$

For this form of the complex number $Z$, the expression of generalized Newon's binomial simplifies considerably, taking into account the properties of the matrices $u_{0}$.

$$[Z]^{l}\equiv[z\,u_{0}+\sigma\,u_{1}]^{l}=\sum_{m=0}^{l}\binom{{}_{p}\rho_{q}(l)}{{}_{p}\rho_{q}(m)}(z\,u_{0})^{l-m}(\sigma\,u_{1})^{m}=(z\,u_{0})^{l}+(\sigma\,u_{1})^{l}=z^{l}\,u_{0}+\sigma^{l}\,u_{1} \qquad (4.3)$$

because the only non-zero terms in the sum are those in which $u_{0}$ and $u_{1}$ "do not meet", that is, they do not multiply each other.

Particularly, we have

$$|Z|^2 = Z Z^* = (z\, u_0 + \sigma\, u_1)(z^*\, u_0 + \sigma^*\, u_1) = |z|^2\, u_0 + |\sigma|^2\, u_1 \qquad (4.4)$$

Therefore, the variable $Z$ can be written as a 2 x 2 diagonal matrix

$$Z^m = z^m\, u_0 + \sigma^m\, u_1 = \begin{pmatrix} z^m & 0 \\ 0 & \sigma^m \end{pmatrix} \qquad (4.5)$$

We can apply the same procedure to the generalized hypergeometric function.

According to the above considerations from paper [13], the expression of coherent states with matrix argument will therefore be

$$|z\, u_0 + \sigma\, u_1 \!>\, = \frac{1}{\sqrt{{}_pF_q(|z\, u_0 + \sigma\, u_1|^2)}} \sum_{n=0}^{\infty} \frac{(z\, u_0 + \sigma\, u_1)^n}{\sqrt{{}_p\rho_q(n)}} |n> \qquad (4.6)$$

$$\begin{aligned}
{}_pF_q(|Z|^2) &= \sum_{n=0}^{\infty} c_n (|Z|^2)^n = \sum_{n=0}^{\infty} c_n (|z|^2\, u_0 + |\sigma|^2\, u_1)^n = \\
&= u_0 \sum_{n=0}^{\infty} c_n (|z|^2)^n + u_1 \sum_{n=0}^{\infty} c_n (|\sigma|^2)^n = u_0\, {}_pF_q(|z|^2) + u_1\, {}_pF_q(|\sigma|^2)
\end{aligned} \qquad (4.7)$$

Using the properties of the matrices $u_0$ and $u_1$, i.e. $(u_j)^n = u_j$ and $u_j \cdot u_k = u_k \cdot u_j = 0$, $j \neq k$, this means that the normalization function of coherent states with shifted arguments breaks down into two parts, each depending on one of the variables.

$${}_pF_q(|z\, u_0|^2 + |\sigma\, u_1|^2) = {}_pF_q(|z\, u_0|^2) + {}_pF_q(|\sigma\, u_1|^2) \qquad (4.8)$$

From a mathematical point of view, this relation actually represent a matrix version of Cauchy's functional equation, whose general expression is

$$\mathcal{F}(x+y) = \mathcal{F}(x) + \mathcal{F}(y) \qquad (4.9)$$

In the case we are examining, when the argument of coherent states is of matrix type, these relations will be written as

$$\mathcal{F}(x\, u_0 + y\, u_1) = \mathcal{F}(x\, u_0) + \mathcal{F}(y\, u_1) = \mathcal{F}(x) u_0 + \mathcal{F}(y) u_1 \qquad (4.10)$$

The product of two such functions is

$$\begin{aligned}
\mathcal{F}(x\, u_0 + y\, u_1)\mathcal{G}(\varsigma\, u_0 + \xi\, u_1) &= [\mathcal{F}(x)u_0 + \mathcal{F}(y)u_1][\mathcal{G}(\varsigma)u_0 + \mathcal{G}(\xi)u_1] = \\
&= \mathcal{F}(x)\mathcal{G}(\varsigma)u_0 + \mathcal{F}(y)\mathcal{G}(\xi)u_1
\end{aligned} \qquad (4.11)$$

Since the matrices $u_0$ and $u_1$ are idempotent, they can be removed from the respective functions, so that the above expression becomes

$${}_pF_q(|z\, u_0|^2 + |\sigma\, u_1|^2) = {}_pF_q(|z|^2) u_0 + {}_pF_q(|\sigma|^2) u_1 \qquad (4.12)$$

If we write the expression of the whole coherent states $|zu_0+\sigma u_1>$ as an expansion over the basis vectors $|n>$, we finally obtain

$$|zu_0+\sigma u_1>=\frac{1}{\sqrt{{}_pF_q(|z|^2)u_0+{}_pF_q(|\sigma|^2)u_1}}\left(u_0\sum_{n=0}^{\infty}\frac{z^n}{\sqrt{{}_p\rho_q(n)}}|n>+u_1\sum_{n=0}^{\infty}\frac{\sigma^n}{\sqrt{{}_p\rho_q(n)}}|n>\right)=$$

$$=\frac{\sqrt{{}_pF_q(|z|^2)u_0}}{\sqrt{{}_pF_q(|z|^2)u_0+{}_pF_q(|\sigma|^2)u_1}}|zu_0>+\frac{\sqrt{{}_pF_q(|\sigma|^2)u_1}}{\sqrt{{}_pF_q(|z|^2)u_0+{}_pF_q(|\sigma|^2)u_1}}|\sigma u_1>$$

(4.13)

where the coherent states for the individual variables are

$$|zu_0>=\frac{1}{\sqrt{{}_pF_q(|z|^2)u_0}}\sum_{n=0}^{\infty}\frac{z^n}{\sqrt{{}_p\rho_q(n)}}|n>u_0\ ,\quad |\sigma u_1>=\frac{1}{\sqrt{{}_pF_q(|\sigma|^2)u_1}}\sum_{n=0}^{\infty}\frac{\sigma^n}{\sqrt{{}_p\rho_q(n)}}|n>u_1$$

(4.14)

Since $<zu_0|zu_0>=1$ and $<\sigma u_1|zu_0>=<\sigma|z>u_1\cdot u_0=0$, and so on, the normalization condition reduces to $<zu_0+\sigma u_1|zu_0+\sigma u_1>=1$, as is correct.

The partial conclusion is that these coherent states with argument sum of two terms can be written as a linear combination of coherent states of the separate terms, having the weights

$$\mathcal{P}_{|zu_0|}\equiv\frac{\sqrt{{}_pF_q(|z|^2 u_0)}}{\sqrt{{}_pF_q(|z|^2)u_0+{}_pF_q(|\sigma|^2)u_1}}\ ,\quad \mathcal{P}_{|\sigma u_1|}\equiv\frac{\sqrt{{}_pF_q(|z|^2)u_1}}{\sqrt{{}_pF_q(|z|^2)u_0+{}_pF_q(|\sigma|^2)u_1}} \quad (4.15)$$

In short, we will write

$$|zu_0+\sigma u_1>=\mathcal{P}_{|zu_0|}|zu_0>+\mathcal{P}_{|\sigma u_1|}|\sigma u_1> \quad (4.16)$$

It is not difficult to verify that the coherent states are normalized to unity:

$$<zu_0+\sigma u_1|zu_0+\sigma u_1>=$$
$$=(\mathcal{P}_{|zu_0|}u_0<z|+\mathcal{P}_{|\sigma u_1|}u_1<\sigma|)(\mathcal{P}_{|zu_0|}|z>u_0+\mathcal{P}_{|\sigma u_1|}|\sigma>u_1)=(\mathcal{P}_{|zu_0|})^2+(\mathcal{P}_{|\sigma u_1|})^2=1$$

(4.17)

as well as that their projector is

$$|zu_0+\sigma u_1><zu_0+\sigma u_1|=(\mathcal{P}_{|zu_0|}|zu_0>+\mathcal{P}_{|\sigma u_1|}|\sigma>u_1)(\mathcal{P}_{|zu_0|}<zu_0|+\mathcal{P}_{|\sigma u_1|}<\sigma u_1|)=$$
$$=(\mathcal{P}_{|zu_0|})^2|zu_0><zu_0|+(\mathcal{P}_{|\sigma u_1|})^2|\sigma u_1><\sigma u_1|$$

(4.18)

It is trivial to observe that the limit for $|zu_0+\sigma u_1>$ reduces the coherent states with the shifted matrix argument to the set of coherent states that label by a single variable:

$$\lim_{\sigma\to 0}|zu_0+\sigma u_1>=|z> \quad (4.19)$$

*Particular case* **1- One-dimensional quantum harmonic oscillator**. Considering $p=q=0$ and $\{a_i\}_1^p = \{b_j\}_1^q$, we will have $_0F_0(/;/;x) = \exp(x)$, we will have the following coherent states with matrix argument, normalized to unity:

$$|zu_0 + \sigma u_1> = \frac{1}{\sqrt{e^{|z|^2 u_0} + e^{|\sigma|^2 u_1}}} \left( u_0 \sum_{n=0}^{\infty} \frac{(z)^n}{\sqrt{n!}} |n> + u_1 \sum_{n=0}^{\infty} \frac{(\sigma)^n}{\sqrt{n!}} |n> \right) =$$

$$= \frac{\sqrt{e^{|z|^2 u_0}}}{\sqrt{e^{|z|^2 u_0} + e^{|\sigma|^2 u_1}}} |zu_0> + \frac{\sqrt{e^{|\sigma|^2 u_1}}}{\sqrt{e^{|z|^2 u_0} + e^{|\sigma|^2 u_1}}} |\sigma u_1>$$

(4.20)

In the above expressions, exponentials appear that have a matrix structure as their exponent, with the interesting property

$$e^{|z|^2 u_0} = e^{\begin{pmatrix} |z|^2 & 0 \\ 0 & 0 \end{pmatrix}} = \sum_{n=0}^{\infty} \frac{1}{n!} \begin{pmatrix} |z|^2 & 0 \\ 0 & 0 \end{pmatrix}^n = \sum_{n=0}^{\infty} \frac{1}{n!} \begin{pmatrix} (|z|^2)^n & 0 \\ 0 & 0 \end{pmatrix} = \begin{pmatrix} \sum_{n=0}^{\infty} \frac{(|z|^2)^n}{n!} & 0 \\ 0 & 0 \end{pmatrix} = \begin{pmatrix} e^{|z|^2} & 0 \\ 0 & 0 \end{pmatrix} \quad (4.21)$$

and similar for the second variable.

Taking the logarithm of these relations, the both matrix being positive-defined matrix, we obtain

$$\begin{pmatrix} |z|^2 & 0 \\ 0 & 0 \end{pmatrix} = \log \begin{pmatrix} e^{|z|^2} & 0 \\ 0 & 0 \end{pmatrix}, \quad \begin{pmatrix} 0 & 0 \\ 0 & |\sigma|^2 \end{pmatrix} = \log \begin{pmatrix} 0 & 0 \\ 0 & e^{|\sigma|^2} \end{pmatrix}$$

$$\begin{pmatrix} |z|^2 & 0 \\ 0 & |\sigma|^2 \end{pmatrix} = \log \begin{pmatrix} e^{|z|^2} & 0 \\ 0 & e^{|\sigma|^2} \end{pmatrix} = \log \left[ e^{|z|^2} u_0 + e^{|\sigma|^2} u_1 \right]$$

(4.22)

This case corresponds to the one-dimensional quantum harmonic oscillator, respectively to the canonical coherent states.

*Particular case* **2 - A two-level system**. For a two-level system we have

$$|zu_0 + \sigma u_1> = \frac{1}{\sqrt{u_0 {}_pF_q(|z|^2) + u_1 {}_pF_q(|\sigma|^2)}} \left( u_0 \sum_{n=0}^{\infty} \frac{z^n}{\sqrt{{}_p\rho_q(n)}} |n> + u_1 \sum_{n=0}^{\infty} \frac{\sigma^n}{\sqrt{{}_p\rho_q(n)}} |n> \right) \quad (4.23)$$

$$<zu_0 + \sigma u_1| = \frac{1}{\sqrt{u_0 {}_pF_q(|z|^2) + u_1 {}_pF_q(|\sigma|^2)}} \left( u_0 \sum_{n=0}^{\infty} \frac{(z^*)^n}{\sqrt{{}_p\rho_q(n)}} <n| + u_1 \sum_{n=0}^{\infty} \frac{(\sigma^*)^n}{\sqrt{{}_p\rho_q(n)}} <n| \right) \quad (4.24)$$

Then, normalization condition is

$$<zu_0 + \sigma u_1 | zu_0 + \sigma u_1> = \frac{1}{u_0 {}_pF_q(|z|^2) + u_1 {}_pF_q(|\sigma|^2)} \left( u_0 \sum_{n=0}^{\infty} \frac{(|z|^2)^n}{{}_p\rho_q(n)} + u_1 \sum_{n=0}^{\infty} \frac{(|\sigma|^2)^n}{{}_p\rho_q(n)} \right) = 1 \quad (4.25)$$

where $_p\rho_q(0) = 1$ and $_p\rho_q(1) = \dfrac{\prod_{j=1}^{q} b_j}{\prod_{i=1}^{p} a_i} \neq 0$.

Coherent states with shifted argument satisfy the completeness relation (or resolution of the identity operator), which is a fundamental property in the construction of the coherent states.

$$\int d\mu(\varepsilon z + \lambda \sigma)|\varepsilon z + \lambda \sigma><\varepsilon z + \lambda \sigma| = 1 \tag{4.26}$$

then this condition is equivalent with

$$\int d\mu(Z)|Z><Z| = 1 \tag{4.27}$$

The expression of the corresponding appropriately integration measure on the integration domain is

$$d\mu(\varepsilon z + \lambda \sigma) = \Gamma(a/b) d(|\varepsilon z + \lambda \sigma|^2) \frac{d\varphi_Z}{2\pi} \times$$
$$\times G_{p,q+1}^{q+1,0}\left(|[\varepsilon z + \lambda \sigma]|^2 \left| \begin{array}{c} /\ ; \\ 0, \{b_j - 1\}_1^q \ ; \end{array} \begin{array}{c} \{a_i - 1\}_1^p \\ / \end{array} \right.\right)_p F_q\left(|[\varepsilon z + \lambda \sigma]|^2\right) \tag{4.28}$$

Let's see how this condition looks in the case of coherent states with matrix argument.

$$\int d\mu(z u_0; \sigma u_1)|z u_0 + \sigma u_1><z u_0 + \sigma u_1| =$$
$$= \int d\mu(z u_0 + \sigma u_1)\left[(P_{|zu_0|})^2 |z u_0><z u_0| + (P_{|\sigma u_1|})^2 |\sigma u_1><\sigma u_1|\right] = 1 \tag{4.29}$$

With the substitutions used, the corresponding integration measure has the expression

$$d\mu(z u_0 ; \sigma u_1) = \Gamma(a/b) \frac{d^2(z u_0 + \sigma u_1)}{\pi} \times$$
$$\times G_{p,q+1}^{q+1,0}\left(|z u_0|^2 + |\sigma u_1|^2 \left| \begin{array}{c} /\ ; \\ 0, \{b_j - 1\}_1^q \ ; \end{array} \begin{array}{c} \{a_i - 1\}_1^p \\ / \end{array} \right.\right)_p F_q\left(u_0|z|^2 + u_1|\sigma|^2\right) \tag{4.30}$$

The integration measure will be

$$d\mu(z u_0 , \sigma u_1) = \Gamma(a/b) \frac{d^2(z u_0 + \sigma u_1)}{\pi} \times$$
$$\times G_{p,q+1}^{q+1,0}\left(|z u_0|^2 + |\sigma u_1|^2 \left| \begin{array}{c} /\ ; \\ 0, \{b_j - 1\}_1^q \ ; \end{array} \begin{array}{c} \{a_i - 1\}_1^p \\ / \end{array} \right.\right)_p F_q\left(u_0|z|^2 + u_1|\sigma|^2\right) \tag{4.31}$$

or, in variable $Z = z u_0 + \sigma u_1$:

$$d\mu(Z) = \Gamma(a/b) \frac{d^2(Z)}{\pi} G_{p,q+1}^{q+1,0}\left(|Z|^2 \left| \begin{array}{c} /\ ; \\ 0, \{b_j - 1\}_1^q \ ; \end{array} \begin{array}{c} \{a_i - 1\}_1^p \\ / \end{array} \right.\right)_p F_q\left(|Z|^2\right) \tag{4.32}$$

The differential of the sum $Z = z u_0 + \sigma u_1$ is actually the differential of a function of two complex variables $z$ and $\sigma$ and is calculated as follows:

$$\frac{d^2Z}{\pi} \equiv \frac{d^2(zu_0 + \sigma u_1)}{\pi} = \frac{\partial}{\partial z}(zu_0 + \sigma u_1)\frac{d^2z}{\pi} + \frac{\partial}{\partial \sigma}(zu_0 + \sigma u_1)\frac{d^2\sigma}{\pi} =$$
$$= \frac{d^2z}{\pi}u_0 + \frac{d^2\sigma}{\pi}u_1 = \frac{d\varphi}{2\pi}d(|z|^2)u_0 + \frac{d\varphi}{2\pi}d(|\sigma|^2)u_1 \qquad (4.33)$$

Let us first examine Meijer's G function from the expression of the integration measure. We use the integral representation of Meijer G-functions [18]. We have successively

$$G_{p,q+1}^{q+1,0}\left(|Z|^2 \left| \begin{array}{c} /\,; \quad \{a_i-1\}_1^p \\ 0,\{b_j-1\}_1^q\,; \quad / \end{array}\right.\right) = \frac{1}{2\pi i}\int_{\mathcal{L}} \frac{\Gamma(-s)\prod_{j=1}^{q}\Gamma(b_j-1-s)}{\prod_{i=1}^{p}\Gamma(a_i-1-s)}(|Z|^2)^s ds \equiv \qquad (4.34)$$

$$\equiv \frac{1}{2\pi i}\int_{\mathcal{L}} \Phi(s;\boldsymbol{a-1},\boldsymbol{b-1})(|Z|^2)^s ds$$

$$G_{p,q+1}^{q+1,0}\left(|Z|^2 \left| \begin{array}{c} /\,; \quad \{a_i-1\}_1^p \\ 0,\{b_j-1\}_1^q\,; \quad / \end{array}\right.\right) = G_{p,q+1}^{q+1,0}\left(|zu_0|^2 + |\sigma u_1|^2 \left| \begin{array}{c} /\,; \quad \{a_i-1\}_1^p \\ 0,\{b_j-1\}_1^q\,; \quad / \end{array}\right.\right) =$$

$$= \frac{1}{2\pi i}\int_{\mathcal{L}} \Phi(s;\boldsymbol{a-1},\boldsymbol{b-1})(|zu_0|^2 + |\sigma u_1|^2)^s ds =$$

$$= \frac{1}{2\pi i}\int_{\mathcal{L}} \Phi(s;\boldsymbol{a-1},\boldsymbol{b-1})(|zu_0|^2)^s ds + \frac{1}{2\pi i}\int_{\mathcal{L}} \Phi(s;\boldsymbol{a-1},\boldsymbol{b-1})(|\sigma u_1|^2)^s ds = \qquad (4.35)$$

$$= u_0 \frac{1}{2\pi i}\int_{\mathcal{L}} \Phi(s;\boldsymbol{a-1},\boldsymbol{b-1})(|z|^2)^s ds + u_1 \frac{1}{2\pi i}\int_{\mathcal{L}} \Phi(s;\boldsymbol{a-1},\boldsymbol{b-1})(|\sigma|^2)^s ds =$$

$$= u_0 G_{p,q+1}^{q+1,0}\left(|zu_0|^2 \left| \begin{array}{c} /\,; \quad \{a_i-1\}_1^p \\ 0,\{b_j-1\}_1^q\,; \quad / \end{array}\right.\right) + u_1 G_{p,q+1}^{q+1,0}\left(|\sigma u_1|^2 \left| \begin{array}{c} /\,; \quad \{a_i-1\}_1^p \\ 0,\{b_j-1\}_1^q\,; \quad / \end{array}\right.\right)$$

Consequently, the total Meijer *G*-function can be decomposed into two individual functions, each depending on one of the variables of the sum $|zu_0|^2 + |\sigma u_1|^2$.

After these results, the integration measure becomes

$$d\mu(zu_0,\sigma u_1) = \Gamma(\boldsymbol{a}/\boldsymbol{b})\left[\frac{d\varphi}{2\pi}d(|z|^2)u_0 + \frac{d\varphi}{2\pi}d(|\sigma|^2)u_1\right]\cdot\left[u_{0\,p}F_q(|z|^2) + u_{1\,p}F_q(|\sigma|^2)\right]\times$$

$$\times\left[u_0 G_{p,q+1}^{q+1,0}\left(|z|^2 \left| \begin{array}{c} /\,; \quad \{a_i-1\}_1^p \\ 0,\{b_j-1\}_1^q\,; \quad / \end{array}\right.\right) + u_1 G_{p,q+1}^{q+1,0}\left(|\sigma|^2 \left| \begin{array}{c} /\,; \quad \{a_i-1\}_1^p \\ 0,\{b_j-1\}_1^q\,; \quad / \end{array}\right.\right)\right] \qquad (4.36)$$

In short, we will be able to write symbolically

$$d\mu(zu_0,\sigma u_1) = \Gamma(\boldsymbol{a}/\boldsymbol{b})[d\mu(z)u_0 + d\mu(\sigma)u_1] \qquad (4.37)$$

so that the decomposition relation of the unit operator becomes

$$\Gamma(a/b)\left[\int d\mu(z)(\mathcal{P}_{|zu_0|})^2 |zu_0\rangle\langle zu_0|\right]u_0 + \Gamma(a/b)\left[\int d\mu(\sigma)(\mathcal{P}_{|\sigma u_1|})^2 |\sigma u_1\rangle\langle \sigma u_1|\right]u_1 = 1 \quad (4.38)$$

The projector of coherent states $|zu_0\rangle\langle zu_0| = u_0|z\rangle\langle z|$ is

$$u_0|z\rangle\langle z| = u_0 \frac{1}{{}_pF_q(|z|^2)} \sum_{n=0}^{\infty} \frac{(|z|^2)^n}{{}_p\rho_q(n)} |n\rangle\langle n| \quad (4.39)$$

Let's integrate the expressions in the right brackets one by one.

$$\Gamma(a/b)\left[\int d\mu(z)(\mathcal{P}_{|zu_0|})^2 |z\rangle\langle z|\right]u_0 =$$

$$= \Gamma(a/b)\left[\sum_{n=0}^{\infty} \frac{|n\rangle\langle n|}{{}_p\rho_q(n)} \int_0^{2\pi} \frac{d\varphi_z}{2\pi} \int_0^{\infty} d(|z|^2)(|z|^2)^n G_{p,q+1}^{q+1,0}\left(|z|^2 \left| \begin{array}{c} /\,; \quad \{a_i-1\}_1^p \\ 0, \{b_j-1\}_1^q \,; \quad / \end{array}\right.\right)\right]u_0 = \quad (4.40)$$

$$= \Gamma(a/b)\left[\sum_{n=0}^{\infty} \frac{|n\rangle\langle n|}{{}_p\rho_q(n)} \frac{1}{\Gamma(a/b)} {}_p\rho_q(n)\right]u_0 = u_0 \sum_{n=0}^{\infty} |n\rangle\langle n| = u_0$$

and similar for the integral over variable $\sigma$, whose result is $u_1$.

$$u_0 + u_1 = \begin{pmatrix} 1 & 0 \\ 0 & 0 \end{pmatrix} + \begin{pmatrix} 0 & 0 \\ 0 & 1 \end{pmatrix} = \begin{pmatrix} 1 & 0 \\ 0 & 1 \end{pmatrix} = I \quad (4.41)$$

i.e. the integral is just equal with the identity operator, as we expected.

Considering the above relation, the sum of the two matrices is the unity matrix itself, it means that the expression of the integration measure is correct, that is, the coherent states having the matrix argument satisfy the resolution of the identity operator. This is the third condition imposed on coherent states in the so-called "Klauder's prescriptions" [19].

The first two conditions in this set are:

1.) The coherent states are normalized, but not orthogonal and form an overcomplete set:

$$\langle zu_0 + \sigma u_1 | z'u_0 + \sigma'u_1 \rangle = \begin{cases} 1, & z = z', \quad \sigma = \sigma', \quad \text{normalization} \\ 0, & z \neq z', \quad \sigma \neq \sigma', \quad \text{non orthogonality} \end{cases} \quad (4.42)$$

The coherent states are continuous in the complex label variable

$$\lim_{z \to z'} \| z - z' \| = \lim_{\sigma \to \sigma'} \| \sigma - \sigma' \| = 0 \quad (4.43)$$

The expected value of an observable $\mathcal{A}$, which characterizes the examined system, in the representation of coherent states is

$$\langle \mathcal{A} \rangle_{z,\sigma} \equiv \langle zu_0 + \sigma u_1 | \mathcal{A} | zu_0 + \sigma u_1 \rangle =$$

$$= \frac{1}{{}_pF_q(|zu_0|^2) + {}_pF_q(|\sigma u_1|^2)} \sum_{n,n'=0}^{\infty} \frac{(zu_0)^{n+n'} + (\sigma u_1)^{n+n'}}{\sqrt{{}_p\rho_q(n){}_p\rho_q(n')}} \langle n|\mathcal{A}|n'\rangle \quad (4.44)$$

If the observable has a diagonal spectrum based on Fock vectors, $\mathcal{A}|n> = A(n)|n>$, and the expression of the eigenvalues can be expanded in power series in $n$, i.e. $A(n) = \sum_j c_j n^j$, the relation becomes

$$\begin{aligned}
<\mathcal{A}>_{z,\sigma} &= \frac{1}{u_0 \,_pF_q(|z|^2) + u_1 \,_pF_q(|\sigma|^2)} \sum_{n=0}^{\infty} \frac{u_0(|z|^2)^n + u_1(|\sigma|^2)^n}{_p\rho_q(n)} \sum_j c_j n^j = \\
&= \frac{1}{u_0 \,_pF_q(|z|^2) + u_1 \,_pF_q(|\sigma|^2)} \sum_j c_j \left[ u_0 \sum_{n=0}^{\infty} n^j \frac{(|z|^2)^n}{_p\rho_q(n)} + u_1 \sum_{n=0}^{\infty} n^j \frac{(|\sigma|^2)^n}{_p\rho_q(n)} + \right] = \\
&= \frac{u_0 \sum_j c_j \left(|z|^2 \frac{\partial}{\partial|z|^2}\right)^j {}_pF_q(|z|^2) + u_1 \sum_j c_j \left(|z|^2 \frac{\partial}{\partial|\sigma|^2}\right)^j {}_pF_q(|\sigma|^2)}{u_0 \,_pF_q(|z|^2) + u_1 \,_pF_q(|\sigma|^2)} = \\
&= \frac{1}{u_0 \,_pF_q(|z|^2) + u_1 \,_pF_q(|\sigma|^2)} \left[ u_0 A\left(|z|^2 \frac{\partial}{\partial|z|^2}\right) {}_pF_q(|z|^2) + u_1 A\left(|z|^2 \frac{\partial}{\partial|\sigma|^2}\right) {}_pF_q(|\sigma|^2) \right]
\end{aligned} \quad (4.45)$$

where, the last line was obtained using the substitution $A(n) \to A\left(|z|^2 \frac{\partial}{\partial|z|^2}\right)$.

### 5. Mixed states

Let us return to the case of a quantum system in thermodynamic equilibrium with its environment, whose mixed states are described by the canonical density operator.

$$\rho = \frac{1}{Z(\beta)} \sum_{n=0}^{n_{max}} e^{-\beta E_n} |n><n| \quad (5.1)$$

In connection with coherent states, the density operator is involved in two types of (quasi) distributions: the Husimi $Q$-distribution function and the quasi $P$-distribution function that appears in the diagonal expansion of the density operator.

The Husimi $Q$-distribution function is defined as the expectation value of the density operator in the coherent states representation:

$$\begin{aligned}
Q(|zu_0 + \sigma u_1|^2) &= <zu_0 + \sigma u_1 | \rho | zu_0 + \sigma u_1> = \\
&= \frac{1}{Z(\beta)} \sum_{n=0}^{n_{max}} e^{-\beta E_n} <zu_0 + \sigma u_1 | n><n | zu_0 + \sigma u_1> = \\
&= \frac{1}{Z(\beta)} \frac{1}{_pF_q(|zu_0|^2) + _pF_q(|\sigma u_1|^2)} \sum_{n=0}^{n_{max}} e^{-\beta E_n} \frac{(|zu_0 + \sigma u_1|^2)^n}{_p\rho_q(n)} = \\
&= \frac{1}{Z(\beta)} \frac{1}{_pF_q(|zu_0|^2) + _pF_q(|\sigma u_1|^2)} \sum_{n=0}^{n_{max}} e^{-\beta E_n} \left[ u_0 \frac{(|z|^2)^n}{_p\rho_q(n)} + u_1 \frac{(|\sigma|^2)^n}{_p\rho_q(n)} \right]
\end{aligned} \quad (5.2)$$

Finally, we obtain

$$Q(|z|^2 u_0 + |\sigma|^2 u_1) = (\mathcal{P}_{|zu_0|})^2 Q(|z|^2) + (\mathcal{P}_{|\sigma u_1|})^2 Q(|\sigma|^2) \quad (5.3)$$

where we used the notation

$$Q(|z|^2) = \frac{1}{{}_pF_q(|z|^2)} \sum_{n=0}^{n_{max}} e^{-\beta E_n} \frac{(|z|^2)^n}{{}_p\rho_q(n)} \tag{5.4}$$

and similar for variable $\sigma$. This expression is nothing more than Husini's Q-function expression for coherent states labeled by a single variable $z$.

The Husimi *Q*-distribution function is always a positive function.

On the one hand, the expansion of the density operator over the Fock's projectors is

$$\rho = \sum_{n=0}^{n_{max}} \frac{e^{-\beta E_n}}{Z(\beta)} |n><n| \tag{5.5}$$

On the other hand, the quasi *P*-distribution function appear as a weight function in the so called diagonal integral expansion of the density operator over the coherent states projectors:

$$\rho = \int d\mu(Z) \, {}_pP_q(|Z|^2) |Z><Z| = \sum_{n=0}^{n_{max}} \frac{|n><n|}{{}_p\rho_q(n)} \int d\mu(Z) \, {}_pP_q(|Z|^2)(|Z|^2)^n \tag{5.6}$$

Equating these expressions we obtain that the integral must have the following value

$$\int d\mu(Z) \, {}_pP_q(|Z|^2)(|Z|^2)^n = \frac{e^{-\beta E_n}}{Z(\beta)} \, {}_p\rho_q(n) \tag{5.7}$$

$$\int \frac{d^2(Z)}{\pi} G^{q+1,0}_{p,\,q+1}\left(|Z|^2 \,\Bigg|\, \begin{matrix} /\,; & \{a_i - 1\}_1^p \\ 0, \{b_j - 1\}_1^q\,; & / \end{matrix}\right) {}_pF_q(|Z|^2) \, {}_pP_q(|Z|^2)(|Z|^2)^n = \frac{1}{\Gamma(a/\ell)} \frac{e^{-\beta E_n}}{Z(\beta)} \, {}_p\rho_q(n) \tag{5.8}$$

Moving on to the variables $z$ and $\sigma$, the above is written as

$$\rho = \int d\mu(z\,u_0; \sigma\,u_1) \, {}_pP_q(|z|^2\,u_0 + |\sigma|^2\,u_1)|z\,u_0 + \sigma\,u_1><z\,u_0 + \sigma\,u_1| \tag{5.9}$$

This expression must be equal to the expression of the density operator expanded onto the Fock vector projectors $|n><n|$.

$$\frac{1}{Z(\beta)} \sum_{n=0}^{n_{max}} e^{-\beta E_n} |n><n| = \int d\mu(z\,u_0; \sigma\,u_1) \, {}_pP_q(|z|^2\,u_0 + |\sigma|^2\,u_1)|z\,u_0 + \sigma\,u_1><z\,u_0 + \sigma\,u_1| \tag{5.10}$$

Writing the coherent state projector in the form

$$|z\,u_0 + \sigma\,u_1><z\,u_0 + \sigma\,u_1| = \frac{1}{{}_pF_q(|z\,u_0 + \sigma\,u_1|^2)} \sum_{n=0}^{\infty} \frac{[|z|^2\,u_0 + |\sigma|^2\,u_1]^n}{{}_p\rho_q(n)} |n><n| \tag{5.11}$$

we obtain

$$\left[\frac{1}{Z(\beta)} \sum_{n=0}^{n_{max}} e^{-\beta E_n}\right] |n><n| = \left[\int d\mu(z\,u_0; \sigma\,u_1) \frac{{}_pP_q(|z|^2\,u_0 + |\sigma|^2\,u_1)}{{}_pF_q(|z\,u_0 + \sigma\,u_1|^2)} \frac{[|z|^2\,u_0 + |\sigma|^2\,u_1]^n}{{}_p\rho_q(n)}\right] |n><n| \tag{5.12}$$

From here it is seen that the equality is satisfied if we have

$$\int d\mu(z\,u_0;\sigma\,u_1)\frac{{}_pP_q(|z|^2\,u_0+|\sigma|^2\,u_1)}{{}_pF_q(|z\,u_0+\sigma\,u_1|^2)}\big[|z|^2\,u_0+|\sigma|^2\,u_1\big]^n=\frac{1}{Z(\beta)}e^{-\beta E_n}\,{}_pP_q(n) \qquad (5.13)$$

Replacing the integration measure and the coherent states with the corresponding expressions, we will obtain, successively:

$$\rho=\Gamma(a/b)\int\left[\frac{d\varphi}{2\pi}d(|z|^2)u_0+\frac{d\varphi}{2\pi}d(|\sigma|^2)u_1\right]\cdot\left[u_0\,{}_pF_q(|z|^2)+u_1\,{}_pF_q(|\sigma|^2)\right]{}_pP_q(|z|^2\,u_0+|\sigma|^2\,u_1)\times$$

$$\times\left[u_0\,G_{p,q+1}^{q+1,0}\!\left(|z|^2\left|\begin{array}{c}/\,;\ \{a_i-1\}_1^p\\0,\{b_j-1\}_1^q\,;\quad/\end{array}\right.\right)+u_1\,G_{p,q+1}^{q+1,0}\!\left(|\sigma|^2\left|\begin{array}{c}/\,;\ \{a_i-1\}_1^p\\0,\{b_j-1\}_1^q\,;\quad/\end{array}\right.\right)\right]\times$$

$$\times\left[u_0(\mathcal{P}_{|z\,u_0|})^2|z><z|+u_1(\mathcal{P}_{|\sigma\,u_1|})^2|\sigma><\sigma|\right]$$

(5.14)

The properties of singular matrices $u_j$ also allow us to decompose the function ${}_pP_q(|z|^2\,u_0+|\sigma|^2\,u_1)$ into a sum that depends on only one variable:

$${}_pP_q(|z|^2\,u_0+|\sigma|^2\,u_1)=u_0\,{}_pP_q(|z|^2)+u_1\,{}_pP_q(|\sigma|^2) \qquad (5.15)$$

Performing the multiplication and taking into account that $u_j\cdot u_k=0$ for $j\neq k$, we will obtain (because angular integrals are equal to 1):

$$\rho=u_0\,\Gamma(a/b)\int_0^\infty d(|z|^2)\,{}_pF_q(|z|^2)\,{}_pP_q(|z|^2)G_{p,q+1}^{q+1,0}\!\left(|z|^2\left|\begin{array}{c}/\,;\ \{a_i-1\}_1^p\\0,\{b_j-1\}_1^q\,;\quad/\end{array}\right.\right)|z><z|+$$

$$+u_1\,\Gamma(a/b)\int_0^\infty d(|\sigma|^2)\,{}_pF_q(|\sigma|^2)\,{}_pP_q(|\sigma|^2)G_{p,q+1}^{q+1,0}\!\left(|\sigma|^2\left|\begin{array}{c}/\,;\ \{a_i-1\}_1^p\\0,\{b_j-1\}_1^q\,;\quad/\end{array}\right.\right)|\sigma><\sigma|$$

(5.16)

It follows that the density operator can be written as a sum of two terms

$$\rho\equiv u_0\,\rho_z+u_1\,\rho_\sigma \qquad (5.17)$$

The variables $z$ and $\sigma$ are equally important in the right-hand side and consequently their expressions will be similar.

Let's evaluate one of the integrals. First let's equate the two expressions for the density operator.

$$\rho_z=\Gamma(a/b)\int_0^\infty d(|z|^2)\,{}_pF_q(|z|^2)\,{}_pP_q(|z|^2)G_{p,q+1}^{q+1,0}\!\left(|z|^2\left|\begin{array}{c}/\,;\ \{a_i-1\}_1^p\\0,\{b_j-1\}_1^q\,;\quad/\end{array}\right.\right)|z><z| \qquad (5.18)$$

which leads to

$$\rho_z = \Gamma(\boldsymbol{a}/\boldsymbol{b})\sum_n \frac{|n><n|}{{}_p\rho_q(n)}\int_0^\infty d(|z|^2)\,{}_pP_q(|z|^2)G_{p,q+1}^{q+1,0}\left(|z|^2 \left| \begin{array}{c} /\;; \quad \{a_i-1\}_1^p \\ 0,\{b_j-1\}_1^q\;; \quad / \end{array}\right.\right)(|z|^2)^n \qquad (5.19)$$

Similar to Eq. (5.13) we will have

$$\int_0^\infty d(|z|^2)\,{}_pP_q(|z|^2)G_{p,q+1}^{q+1,0}\left(|z|^2 \left| \begin{array}{c} /\;; \quad \{a_i-1\}_1^p \\ 0,\{b_j-1\}_1^q\;; \quad / \end{array}\right.\right)(|z|^2)^n = \frac{1}{\Gamma(\boldsymbol{a}/\boldsymbol{b})}\frac{1}{Z(\beta)}e^{-\beta E_n}\,{}_p\rho_q(n) \quad (5.20)$$

Because ${}_p\rho_q(n)$ is composed by Euler gamma functions (see, Eq. (3.7)), this is a Stieltjes moment problem [20].

Generally, the final solution cannot be found unless one knows how the energy eigenvalues depend on the energy quantum number $n$, i.e. $E_n = f(n)$. If this dependence is not linear, the solution is difficult, even impossible to obtain. However, in the case of a quadratic dependence (for example, such as the Morse oscillator), some specific procedures can be found [21].

Let's take *the case of a system with a linear energy spectrum*, with energy eigenvalues $E_n = \hbar\omega n + E_0$. In this case, we have

$$\frac{1}{Z(\beta)}e^{-\beta E_n} = \left(1-e^{-\beta\hbar\omega}\right)\left(e^{-\beta\hbar\omega}\right)^n \quad,\quad Z(\beta) = \sum_{n=0}^\infty e^{-\beta E_n} = e^{-\beta E_0}\sum_{n=0}^\infty \left(e^{-\beta\hbar\omega}\right)^n = \frac{e^{-\beta E_0}}{1-e^{-\beta\hbar\omega}} \qquad (5.21)$$

The last integral equation then becomes

$$\int_0^\infty d(|z|^2)P(|z|^2)G_{p,q+1}^{q+1,0}\left(|z|^2 \left| \begin{array}{c} /\;; \quad \{a_i-1\}_1^p \\ 0,\{b_j-1\}_1^q\;; \quad / \end{array}\right.\right)(|z|^2)^n = $$

$$= \left(1-e^{-\beta\hbar\omega}\right)\left(e^{-\beta\hbar\omega}\right)^n \Gamma(n+1)\frac{\prod_{j=1}^q \Gamma(b_j+n)}{\prod_{i=1}^p \Gamma(a_i+n)} \qquad (5.22)$$

The solution algorithm involves changing the index $n = s-1$ and the solution is the inverse Mellin transform [22].

$$\int_0^\infty d(|z|^2)P(|z|^2)G_{p,q+1}^{q+1,0}\left(|z|^2 \left| \begin{array}{c} /\;; \quad \{a_i-1\}_1^p \\ 0,\{b_j-1\}_1^q\;; \quad / \end{array}\right.\right)(|z|^2)^{s-1} = $$

$$= \left(e^{\beta\hbar\omega}-1\right)\frac{1}{\left(e^{\beta\hbar\omega}\right)^s}\Gamma(s)\frac{\prod_{j=1}^q \Gamma(b_j-1+s)}{\prod_{i=1}^p \Gamma(a_i-1+s)} \qquad (5.23)$$

Finally it is found that the *P*-quasi distribution function for a system with linear energy eigenvalues is [23]

$$_pP_q(|z|^2) = (e^{\beta \hbar \omega} - 1) \frac{G_{p,q+1}^{q+1,0}\left(e^{\beta \hbar \omega}|z|^2 \middle| \begin{array}{cc} /\,; & \{a_i - 1\}_1^p \\ 0, \{b_j - 1\}_1^q\,; & / \end{array}\right)}{G_{p,q+1}^{q+1,0}\left(|z|^2 \middle| \begin{array}{cc} /\,; & \{a_i - 1\}_1^p \\ 0, \{b_j - 1\}_1^q\,; & / \end{array}\right)} \tag{5.24}$$

Obviously, a similar expression will be obtained for the variable $\sigma$.

The thermal expectation value of an observable $\mathcal{A}$ in a mixed state characterized by the density operator $\rho$ is

$$<\mathcal{A}> = \mathrm{Tr}(\rho\mathcal{A}) = u_0\,\mathrm{Tr}(\rho_0\mathcal{A}) + u_1\,\mathrm{Tr}(\rho_1\mathcal{A}) \tag{5.25}$$

Let us apply the results obtained above to *the case of the one-dimensional harmonic oscillator*, with only two energy levels, for which the energy eigenvalues are

$$E_n = \hbar \omega\, e(n) \equiv \hbar \omega \left(n + \frac{1}{2}\right) \quad,\quad E_0 = \frac{1}{2}\hbar\omega \quad,\quad E_1 = \frac{3}{2}\hbar\omega \tag{5.26}$$

From the expression for dimensionless energy eigenvalues $e(n)$ the coefficients involved in the expression of the hypergeometric function will result:

$$_pe_q(n) = n\frac{\prod_{j=1}^{q}(b_j - 1 + n)}{\prod_{i=1}^{p}(a_i - 1 + n)} \quad \Rightarrow \quad p=1\,,\ q=1\,,\ a_1=1\,,\ b_1=\frac{3}{2} \tag{5.27}$$

$$_p\rho_q(n) = \Gamma(a/b)\Gamma(n+1)\frac{\prod_{j=1}^{q}\Gamma(b_j + n)}{\prod_{i=1}^{p}\Gamma(a_i + n)}\,,\ _1\rho_1(n) = \frac{2}{\sqrt{\pi}}\Gamma\!\left(\frac{3}{2} + n\right),\ _1\rho_1(0)=1\,,\ _1\rho_1(1)=\frac{3}{2} \tag{5.28}$$

We will start from the expression of coherent states using the variable $Z$.

$$|Z> = \frac{1}{\sqrt{_1F_1(|Z|^2)}}\sum_{n=0}^{1}\frac{Z^n}{\sqrt{_1\rho_1(n)}}|n><n|\,,\ _1F_1(|Z|^2) = \sum_{n=0}^{1}\frac{(1)_n}{\left(\frac{3}{2}\right)_n}\frac{(|Z|^2)^n}{n!} = 1 + \frac{2}{3}|Z|^2 \tag{5.29}$$

Consequently, the coherent states becomes

$$|Z> = \frac{1}{\sqrt{1+\frac{2}{3}|Z|^2}}\begin{pmatrix}1\\0\end{pmatrix} + \frac{\sqrt{\frac{2}{3}}Z}{\sqrt{1+\frac{2}{3}|Z|^2}}\begin{pmatrix}0\\1\end{pmatrix} \tag{5.30}$$

respectively the projector on the coherent states is

$$|Z><Z| = \frac{1}{1+\frac{2}{3}|Z|^2}u_0 + \frac{\frac{2}{3}|Z|^2}{1+\frac{2}{3}|Z|^2}u_1 \tag{5.31}$$

On the one hand, the density operator is

$$\rho = \sum_{n=0}^{1} \frac{e^{-\beta\hbar\omega\left(n+\frac{1}{2}\right)}}{Z(\beta)} |n\rangle\langle n| = \frac{e^{-\frac{1}{2}\beta\hbar\omega}}{Z(\beta)} u_0 + \frac{e^{-\frac{3}{2}\beta\hbar\omega}}{Z(\beta)} u_1 \quad ; \quad Z(\beta) = e^{-\frac{1}{2}\beta\hbar\omega} + e^{-\frac{3}{2}\beta\hbar\omega} \tag{5.32}$$

and on the other hand, the integral expansion of the density operator over the coherent states projectors is

$$\rho = \int d\mu(Z)\,_1P_1(|Z|^2)|Z\rangle\langle Z| = \sum_{n=0}^{1} \frac{|n\rangle\langle n|}{_1\rho_1(n)} \int d\mu(Z)\,_1P_1(|Z|^2)(|Z|^2)^n \tag{5.33}$$

Consequently, the corresponding integral is

$$\int d\mu(Z)\,_1P_1(|Z|^2)(|Z|^2)^n = \frac{e^{-\beta\hbar\omega\left(n+\frac{1}{2}\right)}}{Z(\beta)}\,_1\rho_1(n) = \frac{2}{\sqrt{\pi}} \frac{e^{-\frac{1}{2}\beta\hbar\omega}}{Z(\beta)} \frac{1}{\left(e^{\beta\hbar\omega}\right)^n} \Gamma\left(\frac{3}{2}+n\right) \tag{5.34}$$

Substituting this result into the density operator expression, leads to

$$\rho = \frac{2}{\sqrt{\pi}} \frac{e^{-\frac{1}{2}\beta\hbar\omega}}{Z(\beta)} \sum_{n=0}^{1} \frac{|n\rangle\langle n|}{_1\rho_1(n)} \left(e^{-\beta\hbar\omega}\right)^n \Gamma\left(\frac{3}{2}+n\right) =$$

$$= \frac{2}{\sqrt{\pi}} \frac{e^{-\frac{1}{2}\beta\hbar\omega}}{Z(\beta)} \left[\frac{1}{2}\Gamma\left(\frac{1}{2}\right) u_0 + \frac{2}{3} e^{-\beta\hbar\omega} \Gamma\left(\frac{5}{2}\right) u_1\right] = \tag{5.35}$$

$$= \frac{e^{-\frac{1}{2}\beta\hbar\omega}}{e^{-\frac{1}{2}\beta\hbar\omega} + e^{-\frac{3}{2}\beta\hbar\omega}} u_0 + \frac{e^{-\frac{3}{2}\beta\hbar\omega}}{e^{-\frac{1}{2}\beta\hbar\omega} + e^{-\frac{3}{2}\beta\hbar\omega}} u_1$$

This result shows that the two expressions of the density operator, the linear expansion onto the projectors of the Fock orthogonal basis, respectively the integral expansion onto the projectors of the coherent states, are equivalent.

### 6. Some applications

Coherent states constitute the specific bridge between classical and quantum mechanics, contributing to the understanding of the classical limit of quantum mechanics (which was actually Schrodinger's starting point in 1926). Among other applications of coherent states (in quantum optics, condensed matter physics, chemistry, radar theory, mathematical physics and so on), it is also to be connected to quantum information and quantum computing, by linking them with the notion of qubit [24], [25], [26].

Because the coherent states play the role of logical qubits in quantum computing [27], at the end of this paper we will show how the coherent states with the complex square singular matrix argument can serve as a qubit.

As we saw previously, a state consistent with the argument specified above has the expression

$$|Z> = \frac{1}{\sqrt{_pF_q(|Z|^2)}} \sum_{n=0}^{n_{max}} \frac{Z^n}{\sqrt{_p\rho_q(n)}} |n> \qquad (6.1)$$

If we examine a two-level system (which is, in fact, a truncation of the above relation, by retaining only the first two terms of the series), this relation becomes

$$|Z> = \frac{1}{\sqrt{_pF_q(|Z|^2)}} \sum_{n=0}^{1} \frac{Z^n}{\sqrt{_p\rho_q(n)}} |n> = \frac{1}{\sqrt{1+\frac{1}{_p\rho_q(1)}|Z|^2}} |0> + \frac{\frac{1}{\sqrt{_p\rho_q(1)}}Z}{\sqrt{1+\frac{1}{_p\rho_q(1)}|Z|^2}} |1> \qquad (6.2)$$

where $_pF_q(|Z|^2) = 1 + \frac{1}{_p\rho_q(1)}|Z|^2$ and $_p\rho_q(1) = \frac{\prod_{j=1}^{q} b_j}{\prod_{i=1}^{p} a_i}$ .

It is observed that the sum of the squares of the complex coefficients of the ground $|0>$ and excited $|1>$ states is equal to 1, so the expression has the form of a qubit.

Returning to matrix notation

$$Z = z\,u_0 + \sigma\,u_1 = \begin{pmatrix} z & 0 \\ 0 & \sigma \end{pmatrix} \qquad (6.3)$$

the expression for the qubit will be written as

$$\left|\begin{pmatrix} z & 0 \\ 0 & \sigma \end{pmatrix}\right> = \frac{1}{\sqrt{1+\frac{1}{_p\rho_q(1)}\begin{pmatrix} |z|^2 & 0 \\ 0 & |\sigma|^2 \end{pmatrix}}} |0> + \frac{\frac{1}{\sqrt{_p\rho_q(1)}}\begin{pmatrix} z & 0 \\ 0 & \sigma \end{pmatrix}}{\sqrt{1+\frac{1}{_p\rho_q(1)}\begin{pmatrix} |z|^2 & 0 \\ 0 & |\sigma|^2 \end{pmatrix}}} |1> \qquad (6.4)$$

Thus we obtained a new mathematical types of qubits, each based on a different physical principle (e.g. photon qubits, thermal qubits, superconducting qubits, spin qubits, and so on). In this context, the type obtained above can be called a *complex matrix qubit*.

Knowing that a single qubit can represent an infinite number of states, corresponding to different pairs of values of the coefficients of states $|0>$ and $|1>$, it is clear that since the 2 x 2 matrix depends on two complex numbers, the representation ability increases infinitely.

From another point of view, the quantum or von Neumann entropy is a very important entity for quantum information and quantum computing [28]. For a mixed state, described by the density operator $\rho$, the von Neumann entropy is defined as

$$S = -\text{Tr}(\rho \log \rho) \qquad (6.5)$$

Let's see what the expression for the von Neumann entropy is for the case when the density operator is expressed by a 2 x 2 diagonal matrix.

$$\rho = \sum_{n=0}^{1} \frac{e^{-\beta E_n}}{Z(\beta)} |n><n| = \begin{pmatrix} \frac{e^{-\beta E_0}}{Z(\beta)} & 0 \\ 0 & \frac{e^{-\beta E_1}}{Z(\beta)} \end{pmatrix} \equiv \begin{pmatrix} w_0 & 0 \\ 0 & w_1 \end{pmatrix} = w_0 \, u_n + w_1 \, u_1 \tag{6.6}$$

$$w_n = \frac{e^{-\beta E_n}}{Z(\beta)} \;,\; Z(\beta) = \sum_{n=0}^{1} e^{-\beta E_n} = e^{-\beta E_0} + e^{-\beta E_1}$$

Having in mind the formula for the trace of operator product

$$\text{Tr}(AB) = \sum_n <n|AB|n> = \sum_n \sum_m <n|A|m><m|B|n> \tag{6.7}$$

as well as $<n|f(\rho)|m> = f(w_n)\delta_{nm}$, we have

$$\text{Tr}(\rho \log \rho) = \sum_{n=0}^{1} <n| \sum_n <n|\rho|n><n|\log \rho|n> \tag{6.8}$$

If the density matrix $\rho$ is sufficient close to the identity matrix $I$, then the logarithm of $\rho$ can be computed by means of the power series (see, e.g. the equation 1.511, page 53 of the Gradshteyn book [29]).

$$\log \rho = \log(I + (\rho - I)) = -\sum_{k=1}^{\infty} \frac{(-1)^k}{k} (\rho - I)^k \tag{6.9}$$

Then, the diagonal matrix elements in the Fock vectors basis is

$$<n|\log \rho|n> = <n|\log(I + (\rho - I))|n> = -\sum_{k=1}^{\infty} \frac{(-1)^k}{k} (<n|\rho|n> - I)^k =$$
$$= -\sum_{k=1}^{\infty} \frac{(-1)^k}{k} (w_n - I)^k = \log w_n \tag{6.10}$$

Then, the von Neumann entropy becomes

$$S = -\text{Tr}(\rho \log \rho) = -\sum_{n=0}^{1} <n|\rho \log \rho|n> =$$
$$= -\sum_{n=0}^{1} <n|\rho|n> \sum_{k=1}^{\infty} \frac{(-1)^k}{k} <n|(\rho - I)^k|n> = -\sum_{n=0}^{1} w_n \sum_{k=1}^{\infty} \frac{(-1)^k}{k} (w_n - I)^k = \tag{6.11}$$
$$= -(w_0 \log w_0 + w_1 \log w_1)$$

Finally, the final expression of the entropy for a mixed state of a two level system is

$$S = -(w_0 \log w_0 + w_1 \log w_1) \tag{6.12}$$

and, because the density probabilities are positive and subunit quantities, $w_0$, $w_1 < 1$, the entropy is positive, as we expected, in analogy with the Shannon entropy.

In matrix language, taking into account that rules of the matrices product, as well as the fact that the operators can be commuted when we calculate the trace of their product, i.e. $\mathrm{Tr}(AB) = \mathrm{Tr}(BA)$, we obtain the same result (see the proof in the Appendix).

**7. Concluding remarks**

A new version of coherent states, examined in this paper, which has as its argument a linear combination of two singular diagonal square matrices and complex variables as coefficients of the linear combination, can be considered, up to a point, to be similar to coherent states with shifted arguments [13]. However, this type of coherent states possesses some specific properties, due to the presence of singular square matrices. The main advantage comes from the fact that most functions (entities) that have as matrix argument the matrix combination $xu_0 + yu_1$ satisfy Cauchy's functional equation, in which the matrices $u_j$, $j = 0, 1$ can be removed from the function argument, in the manner

$$\mathcal{F}(xu_0 + yu_1) = \mathcal{F}(xu_0) + \mathcal{F}(yu_1) = \mathcal{F}(x)u_0 + \mathcal{F}(y)u_1 \tag{7.1}$$

This allows, in most cases, a decoupling of the contribution of the two complex variables that labeled the coherent states, which is not the case for coherent states with shifted (numerical) argument. This new version of coherent states satisfies all the conditions imposed on coherent states, both of pure, as well as the mixed (thermal) states characterized by the density operator. As applications, we examined the connection between these coherent states and the notions of qubits and von Neuman entropy. As a general conclusion, in the paper we showed that coherent states in a matrix sense can also be constructed. As a consequence, the use of this version of coherent states, as application in the field of quantum information has resulted a new type of qubit, which can be called the complex matrix qubit.

Although in the paper we dealt with coherent states labeled by only two variables (this, also for reasons related to the simplicity of writing the formulas), the procedure can also be generalized for the case of $n_{max} < \infty$ complex matrices, in which case singular square matrices of dimensions $n_{max} \times n_{max}$ will be used.

**Appendix**

Let's find the von Neumann entropy formula using matrix language.

$$S = -\mathrm{Tr}(\rho \log \rho) = -\sum_{n=0}^{1} <n|\rho \log \rho|n> = -\sum_{n=0}^{1} <n|\rho|n><n|\log \rho|n> =$$

$$= -\sum_{n=0}^{1} <n|\rho|n><n|\log \rho|n> = -\sum_{n=0}^{1} <n|\begin{pmatrix} w_0 & 0 \\ 0 & w_1 \end{pmatrix}|n><n|\log\begin{pmatrix} w_0 & 0 \\ 0 & w_1 \end{pmatrix}|n> = \tag{A.1}$$

$$= -\sum_{n=0}^{1} <n|\begin{pmatrix} w_0 & 0 \\ 0 & w_1 \end{pmatrix}|n> \sum_{k=1}^{\infty} \frac{(-1)^{k+1}}{k} <n|\left[\begin{pmatrix} w_0 & 0 \\ 0 & w_1 \end{pmatrix} - I\right]^k |n>$$

Due to the following products of matrices

$$< 0 | \begin{pmatrix} w_0 & 0 \\ 0 & w_1 \end{pmatrix} | 0 > = (1 \quad 0) \begin{pmatrix} w_0 & 0 \\ 0 & w_1 \end{pmatrix} \begin{pmatrix} 1 \\ 0 \end{pmatrix} = (w_0 \quad 0) \begin{pmatrix} 1 \\ 0 \end{pmatrix} = w_0$$

$$< 1 | \begin{pmatrix} w_0 & 0 \\ 0 & w_1 \end{pmatrix} | 1 > = (0 \quad 1) \begin{pmatrix} w_0 & 0 \\ 0 & w_1 \end{pmatrix} \begin{pmatrix} 0 \\ 1 \end{pmatrix} = (0 \quad w_1) \begin{pmatrix} 0 \\ 1 \end{pmatrix} = w_1$$
(A.2)

$$\left[ \begin{pmatrix} w_0 & 0 \\ 0 & w_1 \end{pmatrix} - I \right]^k = \begin{pmatrix} w_0 - 1 & 0 \\ 0 & w_1 - 1 \end{pmatrix}^k = \begin{pmatrix} (w_0 - 1)^k & 0 \\ 0 & (w_1 - 1)^k \end{pmatrix}$$
(A.3)

the following matrix elements holds

$$< 0 | \left[ \begin{pmatrix} w_0 & 0 \\ 0 & w_1 \end{pmatrix} - I \right]^k | 0 > = (1 \quad 0) \begin{pmatrix} (w_0 - 1)^k & 0 \\ 0 & (w_1 - 1)^k \end{pmatrix} \begin{pmatrix} 1 \\ 0 \end{pmatrix} = ((w_0 - 1)^k \quad 0) \begin{pmatrix} 1 \\ 0 \end{pmatrix} = (w_0 - 1)^k$$

$$< 1 | \left[ \begin{pmatrix} w_0 & 0 \\ 0 & w_1 \end{pmatrix} - I \right]^k | 1 > = (0 \quad 1) \begin{pmatrix} (w_0 - 1)^k & 0 \\ 0 & (w_1 - 1)^k \end{pmatrix} \begin{pmatrix} 0 \\ 1 \end{pmatrix} = (0 \quad (w_1 - 1)^k) \begin{pmatrix} 0 \\ 1 \end{pmatrix} = (w_1 - 1)^k$$
(A.4)

Then, the entropy expression becomes

$$S = - < 0 | \begin{pmatrix} w_0 & 0 \\ 0 & w_1 \end{pmatrix} | 0 > \sum_{k=1}^{\infty} \frac{(-1)^{k+1}}{k} < 0 | \left[ \begin{pmatrix} w_0 & 0 \\ 0 & w_1 \end{pmatrix} - I \right]^k | 0 > -$$
$$- < 1 | \begin{pmatrix} w_0 & 0 \\ 0 & w_1 \end{pmatrix} | 1 > \sum_{k=1}^{\infty} \frac{(-1)^{k+1}}{k} < 1 | \left[ \begin{pmatrix} w_0 & 0 \\ 0 & w_1 \end{pmatrix} - I \right]^k | 1 > =$$
$$= - w_0 \sum_{k=1}^{\infty} \frac{(-1)^{k+1}}{k} (w_0 - 1)^k - w_1 \sum_{k=1}^{\infty} \frac{(-1)^{k+1}}{k} (w_1 - 1)^k =$$
$$= - (w_0 \log w_0 + w_1 \log w_1)$$
(A.5)

q.e.d

## References

[1] Schrödinger E., *Der stetige Übergang von der Mikro- zur Makromechanik, Die Naturwissenschaften (in German), Springer Science and Business Media LLC.* **14**, 28, 664–666 (1926). http://www.ejtp.com/articles/ejtpv3i11p123.pdf
https://doi.org/10.48550/arXiv.2210.06224

[2] Glauber, Roy J., *Coherent and Incoherent States of the Radiation Field,* Phys. Rev. **131** (6) *2766*–*2788* (1963). doi:10.1103/physrev.131.2766

[3] Klauder J.R., Skagerstam B., *Coherent States: Applications in Physics and Mathematical Physics*, World Scientific, Singapore, 1985.

[4] Perelomov A., *Generalized coherent states and their applications*, Springer, Berlin 1986.

[5] Gazeau J-P., *Coherent States in Quantum Physics*, Wiley-VCH, Berlin, 2009.

[6] Dodonov V. V., *Nonclassical' states in quantum optics: a 'squeezed' review of the first 75 years,* J. Opt. B: Quant. Semiclass. Optics, **4**, 1, R1–R33 (2002). doi*:*10.1088/1464-4266/4/1/201.

[7] Bera S., Das S., Banerjee A., *Bicomplex generalized hypergeometric functions and their applications*, J. Math. Anal. Appl., **550**, 1, 129490 (2025). https://doi.org/10.1016/j.jmaa.2025.129490.

[8] Bera S., Das S., Banerjee A., *On the Bessel function and $n$-dimensional Hankel transform with Bicomplex arguments and coherent states*. 10.48550/arXiv.2507.16973 (2025).

[9] Thirulogasanthar K., Honnouvo G., *Vector coherent states with matrix moment problems,* J. Phys. A: Math. Gen. 37(40) (2003). DOI:10.1088/0305-4470/37/40/014.

[10] Thirulogasanthar K., Honnouvo G., Krzyżak A., *Multi-matrix vector coherent states*, Ann. Phys. (N. Y.), **314**, 1, 119-144 (2004). https://doi.org/10.1016/j.aop.2004.07.006.

[11] Dehghani A., *General displaced* SU*(1, 1) number states: Revisited Available to Purchase,* J. Math. Phys*.* 55, 043502 (2014). https://doi.org/10.1063/1.4868618].

[12] Gazeau J.-P., del Olmo M. A., *SU(1,1)-displaced coherent states, photon counting, and squeezing*, J. Opt. Soc. Am. B 40, 1083-1091 (2023).
 https://opg.optica.org/josab/abstract.cfm?URI=josab-40-5-1083.

[13] Popov D., *Generalized coherent states with shifted (displaced) arguments,* November 2025, DOI:10.48550/arXiv.2511.10285 (2025).

[14] Hamel G., *Eine Basis aller Zahlen und die unstetigen Lösungen der Funktionalgleichung f(x+y)=f(x)+f(y),* Math. Ann. (in German), **60** (3), Leipzig: *459*–462 (1905). doi*:*10.1007/BF01457624*,* S2CID 120063569.

[15] Kuczma M., *An introduction to the theory of functional equations and inequalities. Cauchy's equation and Jensen's inequality.* Basel: Birkhäuser. (2009). ISBN 978-3-7643-8748-8. doi: 10.1007/978-3-7643-8749-5.

[16] Ali S.T., Englis M., Gazeau J-P., *Vector coherent states from Plancherel's theorem, Clifford algebras and matrix domains* J. Phys. A: Math. Gen. 37 6067-6089 (2004).

[17] Aremua I., *Matrix Vector Coherent States for Landau Levels*, Adv. Stud. Theor. Phys., **14,** 6, 237 − 266 (2020). https://doi.org/10.12988/astp.2020.9728].

[18] Beals R., Szmigielski J., *Mijer G-Functions: A gentle Introduction*, Notices of the AMS, **60**, 7, 866-872 (2013).

[19] Klauder J. R., *Continuous-representation theory. I. Postulates of continuous- representation theory*, J. Math. Phys. **4**, 8, 1055-1058 (1963).

[20] Klauder, J. R., Penson, K. A., Sixdeniers, J.-M., *Constructing coherent states through solutions of Stieltjes and Hausdorff moment problems*, Phys. Rev. A, **64**, 1, 013817 (2001). doi: 10.1103/PhysRevA.64.013817.

[21] Popov D., Dong S.-H., Pop N., Sajfert V., Şimon S., *Construction of the Barut–Girardello quasi coherent states for the Morse potential*, Ann. Phys., **339**, 122–134 (2013). https://doi.org/10.1016/j.aop.2013.08.018].

[22] Mathai A. M., Saxena R. K., Generalized Hypergeometric Functions with Applications in Statistics and Physical Sciences, Lect. Notes Math. Vol. 348, Springer-Verlag, Berlin), 1973.

[23] Popov D., Popov M., *Some operatorial properties of the generalized hypergeometric coherent states*, Phys. Scr. **90**, 035101, 1-16 (2015).

[24] Zhang W.-M., Feng D. H., Gilmore R., *Coherent states: Theory and some applications*, Rev. Mod. Phys. 62, 867-927, (1990). DOI: https://doi.org/10.1103/RevModPhys.62.867.

[25] Klauder J. R., Skagerstam Bo-Sture K., *Coherent States: Applications in Physics and Mathematical Physics*, World Scientific, 1985.

[26] Antoine J.-P., Bagarello F., Gazeau J.-P. – Editors, *Coherent States and Their Applications - A Contemporary Panorama*, Springer, 2018.

[27] Popov D., Zaharie I., Sajfert V., Luminosu I., Popov Deian, *Quantum Information in the Frame of Coherent States Representation*. Int. J. Theor. Phys. **47**. 1441-1454 (2008). doi: 10.1007/s10773-007-9586-9.

[28] Nielsen M. A.*,* Chuang I. L., *Quantum Computation and Quantum Information* (10th anniversary ed.). Cambridge: Cambridge Univ. Press, 2010. ISBN 978-0-521-63503-5.

[29] Gradshteyn I. S., Ryshik I. M., *Table of Integrals, Series and Products,* Seventh ed., Academic Press, Amsterdam, 2007.